\documentclass[universe,article,accept,pdftex,moreauthors]{Definitions/mdpi} 
\usepackage{graphicx}
\firstpage{1} 
\pubvolume{1}
\issuenum{1}
\articlenumber{0}
\pubyear{2022}
\copyrightyear{2022}
\datereceived{} 
\dateaccepted{} 
\datepublished{} 
\hreflink{https://doi.org/} 
\Title{Observability of HOFNARs with  SRG/eROSITA }

\TitleCitation{Observability of HOFNARs with  SRG/eROSITA}

\Author{ Alena D.\ Khokhriakova $^{1,2,3}$\orcidA{}, 
 Andrey I. Chugunov $^{4}$\orcidC{}, Sergei B.\ Popov$^{1,2}$*\orcidP{}, 
  Mikhail E.\ Gusakov $^{4}$\orcidG{},
 and Elena M.\ Kantor $^{4}$
 }

\AuthorNames{Alena Khokhriakova, Andrey Chugunov, Sergei Popov, Mikhail Gusakov and Elena Kantor}

\AuthorCitation{Khokhriakova, A.; Chugunov, A.; Popov, S.; Gusakov, M.; Kantor, E.}

\address{%
$^{1}$ \quad Department of Physics, Lomonosov  Moscow State University,  Russia \\
$^{2}$ \quad Sternberg Astronomical Institute,  119234 Moscow, Russia \\
$^{3}$ \quad ``Basis'' foundation fellow, Moscow, Russia\\
$^{4}$ \quad Ioffe Institute, 194021 St. Petersburg, Russia}
\corres{Correspondence:  sergepolar@gmail.com; 
}

\abstract{
Neutron stars can appear as sources of different nature. 
In this paper we address observability of a hypothetical class of neutron stars
-- HOt and Fast Non Accreting Rotators, HOFNARs. These objects are heated due to the r-mode 
instability. With surface temperatures $\sim 10^6$~K they are expected to be thermal soft X-ray 
emitters. We perform a population synthesis modeling of HOFNARs to predict the number of 
potentially detectable sources in the eROSITA all-sky survey. 
For surface temperatures $\sim 10^6$~K we obtain $\sim 500$ sources 
above the detection limit 0.01~cts~s$^{-1}$ and $\sim 100$ easier identifiable sources with $>0.1$~cts~s$^{-1}$. 
Temperatures $\gtrsim 1.2\times 10^6$~K start to be in contradiction with non-detection of HOFNARs by ROSAT.
Only for $T\lesssim 5\times 10^5$~K numbers predicted for eROSITA turn out to be so low that 
identification does not look possible.
We conclude that eROSITA has good chances to discover HOFNARs, if they exist. Non-detection will 
put very stringent limits on the properties of this type of neutron stars.
}

\keyword{neutron stars; magnetic field; millisecond radio pulsars; X-ray binary systems} 

\begin{document}




\section{Introduction}

Neutron stars (NSs) are the final products of evolution of massive stars. These compact objects, formed after a collapse of a stellar core,  contain in their interiors the densest matter in the Universe \cite{ShapiroTeukolsky,hpy07,CompStarBook18}. Extreme internal properties, unusual composition, and complicated structures in different layers of NSs pose many exciting problems in modern astrophysics. Despite small sizes (typical radii $\sim 10$--12~km) NSs reveal themselves in the whole range of the electromagnetic spectrum (e.g., \cite{Harding13}), as  well as in gravitational waves \cite{Abbott_ea19} bringing us to the era of multi-messenger astronomy. Variety of observational manifestations allow to distinguish many different kinds of NSs forming a zoo of peculiar sources associated with a broad range of 
 physical processes, often non-reproducible in Earth laboratories \cite{CompStarBook18}. 

In this paper we consider detectability of a hypothetical class of NSs -- HOFNARs (HOt and Fast Non 
Accreting Rotators), -- which was recently proposed in \cite{cgk14} by eROSITA (extended ROentgen Survey with an Imaging Telescope Array) \cite{2021A&A...647A...1P}. 
Similar to a well known class of the most rapidly rotating  NSs -- millisecond radio pulsars (MSPs, see e.g., \cite{MSPs22}) -- HOFNARs are formed in low-mass X-ray binaries (LMXBs), where a NS is spun-up (``recycled'') by accretion of matter transferred from a Roche-lobe filling companion \cite{bkk76,acrs82}.

In spite of common origin,  observational properties of MSPs and HOFNARs are much different. MSPs reveal themselves mostly as sources of periodic radio or/and {  high energy (X-ray and/or $\gamma$-ray)} signals, which are associated with magnetospheric processes \cite{Harding22}. Due to old ages and limited heating power inside them \cite{Reisenegger97,KG21} they are rather cold stars (e.g., \cite{Durant_ea12,Schwenzer_ea17,cgk17,Bhattacharya_ea17,GGR19,Boztepe_ea20}). Detectable thermal emission is typically associated with a small part of the surface -- hot spots, -- which 
are
supposed to be heated by magnetospheric currents \cite{HM01,HM02}. 

In contrast to MSPs, HOFNARs might have a strong heating source inside \cite{cgk14}. It is a gravitational wave-driven r-mode instability \cite{fs78a,fs78b,andersson98} which effectively converts the rotational energy into gravitational waves and heat (e.g., \cite{Chugunov17}). The r-mode instability keeps HOFNARs hot: the surface temperature has a typical value  $T\sim 10^6$~K, while the internal temperature is $T_\mathrm{int}\sim 10^8$~K.%
{  
\footnote{
  Similar r-mode heating can take place in newly born neutron stars, if they rotates enough rapidly (e.g., \cite{Lindblom:1998wf, 2003pasb.conf..231R, Alford:2012yn, Routray:2021cdx}). However, they likely evolve very fast and quickly leave the r-mode instability region due to higher temperatures and thus, stronger neutrino luminosity.}
}

As argued in \cite{cgk14}, the magnetic field of HOFNARs is likely dissipated due to Ohmic losses, so these objects might not produce radio pulsar emission. Thus,  observational properties of HOFNARs are mainly associated with a thermal emission from the whole surface (with a possible contribution from the 
emission of a companion star which is exhausted by accretion and does not fill the Roche lobe any more). As far as this thermal emission is primarily in the soft X-ray band, the eROSITA all-sky survey becomes the best opportunity to detect these objects
in the Galactic disc (some dim X-ray candidates are already detected in globular clusters, see \cite{cgk14} for discussion).

In this paper we model the HOFNAR population in the Galactic disc, assuming that these sources have the same {  spatial} distribution as MSPs (Sec.~\ref{Sec_Model}). In Sec.~\ref{Sec_Res} we present  calculations of observational properties of HOFNARs, focusing on the eROSITA all-sky survey. 
In Sec.~\ref{Sec_Discus} we {  at first} confront predictions of our model with the ROSAT data and {  then} discuss identification of X-ray candidates and constraints that will be provided by the eROSITA all-sky survey on the HOFNAR population. 

\section{Model} 
\label{Sec_Model}

In this study we use a population synthesis model (see, e.g. a review 
of
this method in \cite{2007PhyU...50.1123P}) to calculate properties of HOFNARs in the Galactic disc 
in application to observations by eROSITA in the survey mode. In the following subsections we 
describe basics of our model. 
This includes: spatial distribution of HOFNARs and their number, their thermal properties, properties of the detector and the survey, interstellar absorption.


\subsection{Spatial distribution and number of objects}

To model the Galactic population of HOFNARs, which are still a hypothetical class of NSs, we should start with a set of assumptions.
Firstly, as far as HOFNARs and MSPs originate from LMXBs  it is reasonable to assume that they should have similar spatial distribution. 
Secondly, the lifetime of HOFNARs can be as large as $\sim 10^{10}$~yr \cite{cgk14}, which is of the same order as the age of the Galaxy and similar to the lifetime of MSPs. Thus, we
accept that HOFNARs are accumulated in the Galaxy, similar to MSPs.
This allows us to assume that the present day spatial distribution of HOFNARs in the Galactic disc is identical to that of MSPs, and the number of HOFNARs is a given fraction (see below) of the number of MSPs.

{  
Strong neutrino emission can lead to a shorter HOFNAR lifetime (see eq.\ (\ref{Evolution}) and \cite{cgk14}) and thus, some discrepancy in the spatial distribution between MSPs and HOFNARs can appear.
Differences in the evolution of HOFNARs and MSPs progenitor systems can also contribute to the discrepancy. However, the general effect of this discrepancy on the final results is expected to be moderate (in comparison with other uncertainties) and accounting for that difference requires exact knowledge of HOFNARs parameters and calculations of their spatial evolution. Thus, at the stage when no objects of this type are securely identified we prefer to leave these subtleties beyond the scope of this paper. }

The MSP spatial distribution was studied in many papers (see, e.g. \cite{2020JCAP...12..035P, 2018ApJ...863..199G} and references therein). In our modeling we take this distribution from \cite{2007ApJ...671..713S} (however, we do not consider objects in the bulge and in globular clusters)
and use a cylindrical coordinate system $(R,\ \phi,\ z)$, where $z$-axis is perpendicular to the Galactic disc and the Galactic center is located at $R=0$, $z=0$.

{  MSP spatial number density $\xi_\mathrm{MSP}$ is described as:
\begin{equation}
\xi_\mathrm{MSP}(R, z)=N_\mathrm{MSP}\xi_\mathrm{z}(z) \xi_\mathrm{R}(R),
\label{MSP}
\end{equation}
where $ N_\mathrm{MSP} $ is the total number of MSPs and the two functions of $z$ and $R$  are described below.  }
We fitted equilibrium distributions from Fig.~1 in \cite{2007ApJ...671..713S} by the following expressions.
For the $z$-distribution:
\begin{equation}
    \xi_\mathrm{z}(z) = 
    \begin{cases}
    A \exp(-|z| / z_1), &\text{if }|z| \leq z_c,\\
    B \exp(-|z| / z_2), &\text{if }|z| > z_c,
    \end{cases}
\end{equation}
where $A$ and $B=A \exp(-z_c/z_1) \exp(z_c/z_2)$ are normalization constants. Other parameters have the following values: $z_1 = 0.4$ kpc, $z_2 = 1$ kpc, and $z_c = 1$ kpc.
The constant $A$ is derived from the normalization condition:
\begin{equation}
    \int_{-\infty}^{\infty} \xi_\mathrm{z}(z) dz = 1.
\end{equation}
In fact, in our code we use just reasonably large values for the integration limits, specifically $ \pm 10$ kpc (since the numbers are {  sufficiently} big the exact value does not influence the results).

For the radial coordinate ($R$) the distribution is the following:
\begin{equation}
   \xi_\mathrm{R}(R) = C \times R \times \exp(-R/R_0),
\end{equation}
where $C$ is the normalization constant and $R_0 = 4.5$ kpc.
The constant $C$ is derived from the condition:
\begin{equation}
    \int_{0}^{\infty} \xi_\mathrm{R}(R) dR = 1.
\end{equation}
In the code for the upper limit we use the value 40 kpc (the exact choice does not affect the result of normalization).

For the angle $\phi$ in the Galactic plane we use a uniform distribution, i.e. the whole distribution has a cylindrical symmetry. 

The distribution described above might be normalized to get the absolute number of HOFNARs in the Galactic disc.
However, it is difficult to predict this normalization theoretically because there are no detailed 
estimates of the fraction of LMXBs which produce HOFNARs (the remaining LMXBs {  with NSs} might mostly produce MSPs).

The main uncertainty is related to unknown properties of instability windows --- the regions in the temperature-spin frequency diagram, where a NS is unstable with respect to the r-modes. The instability windows strongly depend on the NS microphysics, including composition and superfluid properties  (see, e.g.\ \cite{Haskel15} for a review and \cite{Kantor_ea20,Kantor_ea21,Kraav_ea21} for some recent results). Thus,  detection of HOFNARs by eROSITA will provide crucial constraints on internal properties of NSs.
In this study, we apply a fiducial normalization 
that the ratio of HOFNARs to MSPs is $\nu=0.05$.
{  It is worth to note, that $\nu$, as it is applied in our study, should be treated as a present day value, not surely equal to the HOFNAR to MSP ratio at the origin, which can differ, e.g., if  the average HOFNAR life time is lower than the average MSP lifetime.}

{  Justification of } this number is based, {  in the first place,} on observations of globular clusters. %
Namely, X-ray observations of the globular cluster 47 Tucanae have revealed two sources which are classified as 
LMXBs in a quiescent state (qLMXBs), due to their pure thermal X-ray spectra \cite{Heinke_ea03,Bogdanov_ea16}. However, according to the analysis in \cite{cgk14}, observational properties of these sources suggested that they could be considered as  HOFNAR candidates.
Later, three additional qLMXB candidates  were discovered in 47 Tucanae \cite{Heinke_ea05}, which 
again could be treated  as 
HOFNAR candidates.
Assuming that a reasonable part of these X-ray sources are indeed HOFNARs and taking into account 
that 29 MSPs are currently known in this globular cluster, we conclude that  {  an upper limit on the} number of HOFNARs 
should be of order 
{  10-15\%} of the number of currently known  MSPs.
This estimate is also generally consistent
with observations of transiently accreting NSs analyzed in \cite{gck14a, gck14b,cgk17,Kantor_ea21}: 
at least 3 out of 20 among these NSs would become HOFNARs 
for r-mode instability windows, considered in these references, 
if accretion 
were
terminated now.

Note, however, that
a significant number of MSPs can still be undiscovered in globular clusters
(e.g., \cite{Martsen_ea22}), suggesting that the actual value of $\nu$ can be lower {  than the upper limit of 10-15\%}.

Another difficulty can be associated with the LMXB evolution.
In globular clusters it can be much more {  perplexed} 
than in the Galactic disc due to influence of close encounters  (e.g., \cite{Ivanova_etal08}).
This difference in evolution can lead to different {  rates} of HOFNARs formation  in globular clusters and the Galactic disc.

{  Besides that, globular clusters are generally older than the Galactic disc and if the typical HOFNAR lifetime is less that our fiducial value $10^{10}$~yrs, then significant fraction of HOFNARs, which was formed in globular clusters, could already die out (i.e. leave the r-mode instability region and cool down, see the final part of the evolution tracks in Fig.\ 1 of \cite{cgk14}).  In this case,  the present day fraction of HOFNARs in the Galactic disc can be larger than in globular clusters, leading to an additional uncertainty.}
Fortunately, dependence of our results  (in particular $\log N-\log S$ diagram, Fig.\ 
\ref{fig:logNlogS}) on $\nu$  is linear. So, 
our 
results can be easily re-scaled for any value of the normalization.
{  Thus,  in our modeling we apply $\nu=0.05$ as a fiducial value, having in mind that the actual value of $\nu$ should 
be constrained by observations, e.g. 
by
the eROSITA all-sky survey (as we demonstrate, this normalization is consistent with undetection of HOFNARs by ROSAT, see Section \ref{Sec_ROSAT}).
We underline that the uncertainty in $\nu$ is the most important factor in our model. 
} 

The total number of MSPs in the Galaxy (and, separately, in the disc) is rather uncertain (see 
discussion and references, e.g. in \cite{2003ApJ...597.1036P}). 
To normalize the MSP distribution we appeal to the results of Ref.\ \cite{2018ApJ...863..199G}.
Namely, we assume a constant MSP birthrate $\sim 4.5$~per Myr (corresponding to the HOFNAR birth rate {  $\sim 0.225$~Myr$^{-1}$ for $\nu=0.05$}) over the past 10 Gyr. This gives us the expected number of HOFNARs in the Galactic disc {  $N_\mathrm{HOFNAR}=2250$. This number is substituted in eq.~(\ref{MSP}) wich defines $z$ and $R$ distribution of objects of interest.} We use this value of $N_\mathrm{HOFNAR}$ in calculations below. {  Synthetic objects are distributed in the Galaxy randomly according to the described distribution.}

\subsection{Thermal properties of HOFNARs}

The next assumption is related to the temperature distribution of HOFNARs. It is rather uncertain 
because it {  obviously } depends on 
{  the temperature variation in course of the HOFNAR evolution as well as on the NS properties at the preceding (LMXB) stage of evolution:
at the end of the LMXB stage the NS should be r-mode unstable in order to become a HOFNAR. 
This evolution is complicated and depends, e.g., on the shape of the instability window 
(see e.g., \cite{1998PhRvL..80.4843L, Levin:2000vq,2002PhRvD..65f3006L,2006PhRvD..74d4040G,2006MNRAS.371.1311G, 2009MNRAS.397.1464H, 2012PhRvD..85b4007A, 2014PhRvL.113y1102A, 2014MNRAS.441.1662H, Haskel15,Kokkotas:2015gea, Pattnaik:2017ttf, Ofengeim_2019, Ofengeim_ea19_BulkVisc, Kantor_ea21} for a variety of available models),
neutrino cooling efficiency (see  e.g., \cite{cgk17}), binary evolution history (in particular, evolution of the accretion rate \cite{2013ApJ...775...27C}), accretion torque models (e.g., \cite{2016ApJ...822...33P,2017ApJ...835....4B}), distribution of neutron star masses (e.g., \cite{2016ARA&A..54..401O, 2020arXiv201108157H}), binary properties, etc.
Indeed, if models for all of above mentioned  processes are specified then in principle, one can predict temperature distribution of HOFNARs, but the result 
is obviously model dependent and we leave such analysis beyond the scope of the present paper. Instead of that,
 here} we apply the simplest approach and 
assume that all HOFNARs have the same temperature, which is constant during the whole life of a 
HOFNAR. This assumption corresponds to the instability window shown in panel (a) of Fig.\ 1 in 
\cite{cgk14}. 
{  Note, that results (e.g., $\log N-\log S$ diagram) for any more realistic temperature distribution can be easily obtained as a linear combination of our fixed-temperature results with appropriate weights (e.g., via approximation of the temperature distribution by a histogram).} 

We consider several fiducial values of the temperature which are based on observations of qLMXB/HOFNAR candidates in globular clusters.
We use the unredshifted surface temperature $T=10^6$~K for most of illustrations. However, for 
calculations of observable fluxes we take into account the gravitational redshift, which is 
determined by masses and radii of NSs. 

To calculate properly the spectrum measured by a remote observer, we need to specify the mass 
distribution for HOFNARs as well as 
the mass-radius 
relation.
However, as discussed above, our calculations already include some uncertainties in the input 
parameters, 
thus
it seems  unreasonable to analyze in detail the sensitivity of our results 
to
the mass distribution. 
For simplicity, we assume that all HOFNARs have a mass $M=1.4\,M_\odot$ and circumferential radius 
$R_{\mathrm{NS}}=12$~km.

\subsection{Detectability of HOFNARs}

%
To estimate the number of HOFNARs observable by eROSITA, it is necessary 
to calculate 
the total number of photons detectable from a given HOFNAR per unit time (the eROSITA photon count 
rate).
This rate depends on the shape of a HOFNAR spectrum and on properties of the detector,
both are discussed below.
%

\begin{figure}[H]
    \includegraphics[width=\textwidth]{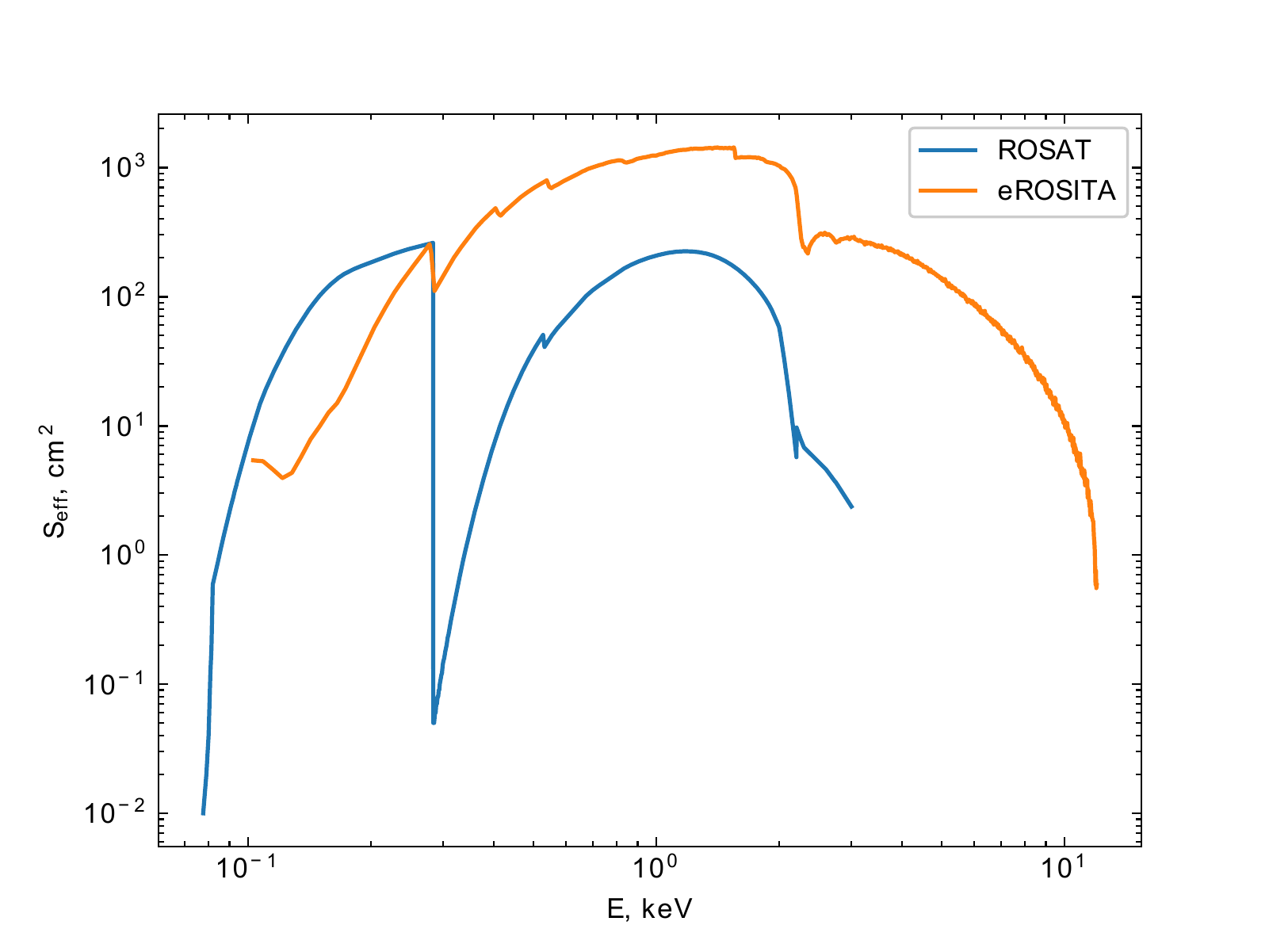}
    \caption{Effective area of eROSITA and ROSAT telescopes versus photon energy. The data for eROSITA are presented under the assumption that signals from ﬁve out of seven telescopes pass through ﬁlters cutting oﬀ the soft part of the spectrum. The data for eROSITA are taken from the site {   \url{https://wiki.mpe.mpg.de/eRosita/erocalib_calibration}}, for ROSAT --- from \cite{1987SPIE..733..519P}.
    \label{fig:Seff}}
\end{figure}

After we fix the surface temperature and redshift, we have to 
specify a NS atmosphere.
We assume that all HOFNARs have  hydrogen atmospheres as expected for 
NSs
in LXMBs. 
For our calculations we apply the {\sc XSPEC} hydrogen atmosphere model {\sc nsatmos} 
\cite{hrng06}. Also, we performed calculations  with the {\sc nsa} \cite{zps96} atmosphere model. The 
difference between the results 
is 
negligible, {  so results for this model are not reported here}.

 eROSITA is an instrument on-board of 
 the Spectrum-Roentgen-Gamma (SRG) mission. It provides the best imaging survey in X-rays
 to date
  \cite{merloni2012, 2021A&A...647A...1P}. The {  working} energy range of  eROSITA is $\sim 0.2$ --- 10 keV 
  (see its effective area in Fig.~\ref{fig:Seff}). The optical system of eROSITA consists of seven 
  elements. Signal from several of them passes through an additional filter which cuts the soft 
  part 
  of the spectrum. We used data on the effective area under the assumption that such filters are 
  used in five out of the seven elements. The corresponding data are taken
from the web-site {  \url{https://wiki.mpe.mpg.de/eRosita/erocalib_calibration}}. 
Initial plans included a 4-year survey (2019-2023), during which eight full scans might be done.
 At the moment the survey is terminated, but it already has completed slightly  more than one half 
 of the planned scans. So, generally, initial goals are reached.
That is why we apply the sensitivity of the complete survey (we also hope that in the near future 
the survey will continue to finalize four final scans).  

To calculate count rates (CR) of the modeled sources we use the following expression:
\begin{equation}
    \mathrm{CR} = \int_{E_1}^{E_2} \frac{N_\mathrm{ph}(E) e^{-\sigma N_H} S_{\mathrm{eff}}(E)}{4 \pi r^2} dE,
    \label{eq:cr}
\end{equation}
where $E$ is the photon energy,
$S_{\mathrm{eff}}$ is the effective area of eROSITA,
$E_1$ and $E_2$ are bounds of the range of sensitivity of the detector,
$r$ is the distance to the source,
$N_\mathrm{H}$ is the column density of hydrogen 
on the line of sight. The flux is reduced due to absorption by the factor $e^{-\sigma 
N_\mathrm{H}}$ (see the next subsection).
To calculate the $\sigma$ value, we use the equation and coefficients {  $C_0$, $C_1$, $C_2$} from \cite{morrison1983}:
\begin{equation}
\sigma=\frac{1}{E}C_2+\frac{1}{E^2}C_1+\frac{1}{E^3}C_0.
\end{equation}
 Finally,
$N_\mathrm{ph}(E) dE$ is the number of photons {  emitted} from the whole surface of an isotropic source in the spectral interval $dE$
per unit time for a distant observer {  (i.e., accounting for general relativistic effects)}.  

We consider a source as potentially visible if its count rate exceeds the eROSITA detectability threshold 0.01~cts~s$^{-1}$. {  Detailed calculations of this limit can be found in \cite{2021ARep...65..615K}. Also, this value corresponds to the limiting sensitivity flux  $1.1 \times 10^{-14}$ erg s$^{-1}$ cm$^{-2}$ in the soft band [0.2 - 2.3] keV  given in \cite{2021A&A...647A...1P} assuming that the spectrum is a power-law with $\Gamma = 1.8$ and no absorption is taken into account. This sensitivity corresponds to the total un-vignetted exposure equal to 1600 s. Effective (vignetted) exposure can be computed by dividing the total exposure by 1.88 in the soft band. }

\subsection{Interstellar absorption}
\label{ism}

Soft X-ray flux can be significantly weakened on the way to the Earth.
To account for effects of propagation  in the interstellar medium we need to specify a 3D distribution of the gas and then to calculate absorption. 

3D distribution of the interstellar medium (separately for different constituents) in the whole Galactic volume is  not well-known (see, e.g. discussion and references in \cite{2016PASJ...68...63S}). Luckily, for our purposes we can neglect some details focusing on the general shape of the distribution.
We proceed as follows.
At first, we adopt distributions of molecular (H$_2$) and atomic (HI) hydrogen according to \cite{2006A&A...459..113M}. 
This provides us with the shape of the 3D distribution of the absorbing medium.
For molecular hydrogen we have:
\begin{equation}
    \rho_{\mathrm{H}_2} (R, z) = \rho_{\mathrm{H}_2}(0,0) \exp\left(-\frac{R}{h_{\mathrm{H}_2}} - \frac{|z|}{z_{\mathrm{H}_2}}\right),
\end{equation}
where $\rho_{\mathrm{H}_2}(0,0) = 4.06$ cm$^{-3}$ is the number density of  H$_2$ molecules at the center of the Galaxy, $h_{\mathrm{H}_2} = 2.57$ kpc is the scale in the radial direction, and $z_{\mathrm{H}_2} = 0.08$ kpc is the scale height of the distribution.

And for atomic hydrogen:
\begin{equation}
    \rho_{\mathrm{HI}} (R, z) = 
    \begin{cases}
    \rho_{\mathrm{HI}}(0, 0) \exp\left(-\frac{R}{h_{\mathrm{HI}}} - \frac{|z|}{z_{\mathrm{HI}}}\right),& \sqrt{R^2 + z^2} > R_t\\
    0,& \sqrt{R^2 + z^2} < R_t.
    \end{cases}
\end{equation}
Then, $\rho_{\mathrm{HI}}(0,0) = 0.32$ cm$^{-3}$ is the HI number density at the center of the Galaxy (i.e., at the center of the untruncated disk), $R_t = 2.75$ kpc is the inner truncation radius, $h_{\mathrm{HI}} = 18.24$ kpc is the radial scale, and $z_{\mathrm{HI}} = 0.52$ kpc is the scaleheight of the distribution.

The total hydrogen column density is obtained by integration along the line of sight. It can be written as:
\begin{equation}
    N_\mathrm{H} = \int (2 \rho_{\mathrm{H}_2} + \rho_{\mathrm{HI}}) dr.
    \label{eq:nh}
\end{equation}
The integral in eq.(\ref{eq:nh}) is calculated numerically.

\begin{figure}[t]
    \includegraphics[width=\textwidth]{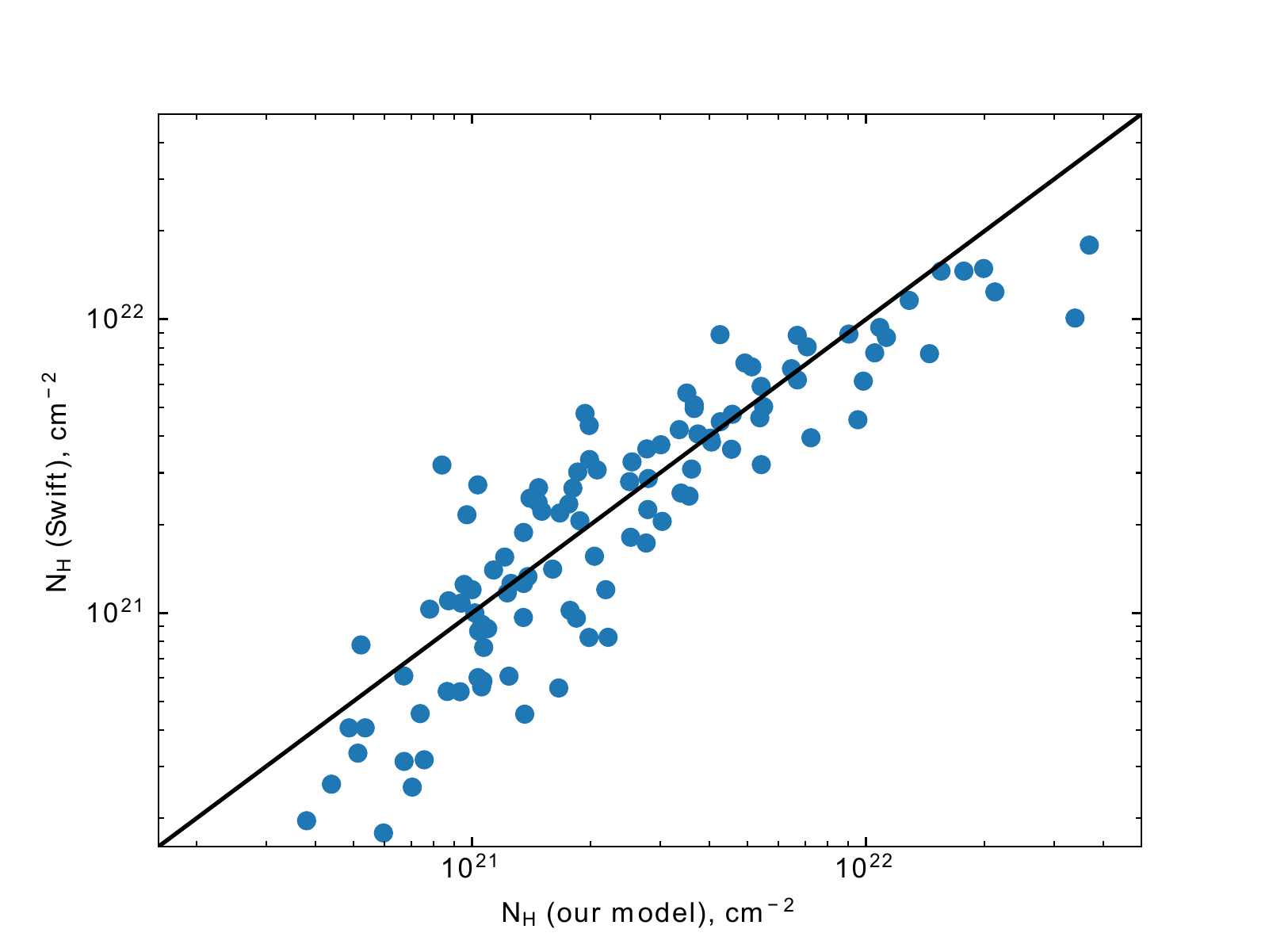}
    \caption{Comparison of hydrogen column density throw the whole Galaxy computed using our model and obtained with the web-tool  \url{https://www.swift.ac.uk/analysis/nhtot/index.php} \cite{2013MNRAS.431..394W}. Solid diagonal line is the bisector. $N_\mathrm{H}$ is shown for 100 directions corresponding to synthetic sources with CR$>0.1$~cts~s$^{-1}$. \label{fig:NH}}
\end{figure}

{  We confront values of $N_\mathrm{H}$ from eq.~(\ref{eq:nh}) integrated up to a very large distance (1 Mpc) with values provided by the on-line service at the Swift web-site in UK (\url{https://www.swift.ac.uk/analysis/nhtot/index.php}) for Galactic $N_\mathrm{H}$ \cite{2013MNRAS.431..394W}. This comparison is shown in  Fig.~\ref{fig:NH}. To select directions in the sky we use coordinates of 100 synthetic sources with count rates $> 0.1$~cts~s$^{-1}$. We find that our model is in good agreement with the data given by the web-tool.}


\section{Results} 
\label{Sec_Res}


\begin{figure}[t]
    \includegraphics[width=\textwidth]{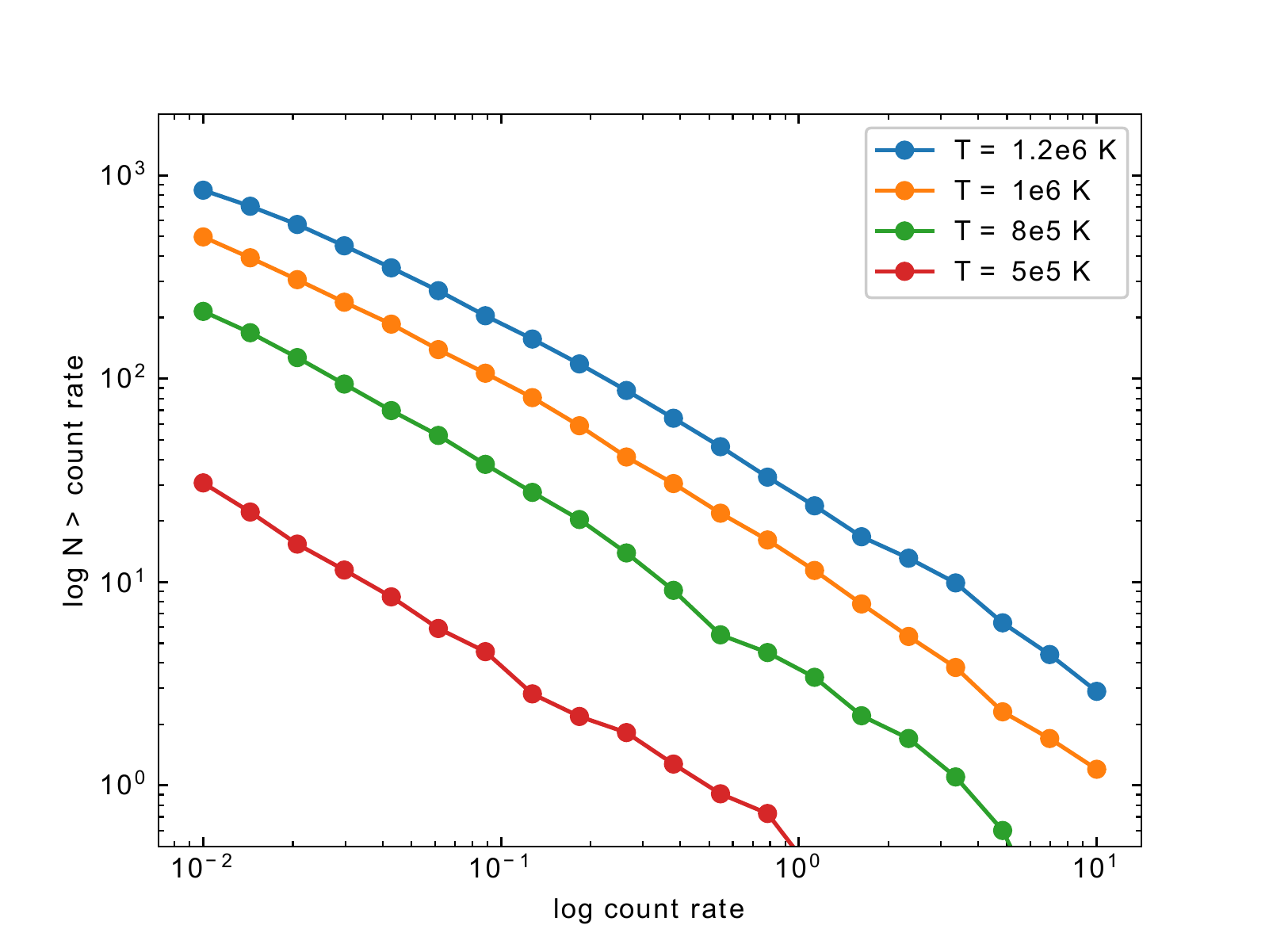}
    \caption{Log $N$ --- Log $ S$ distributions {  (averaged over 10 runs, 2250 objects each)} for different {  unredshifted} surface temperatures of sources. The horizontal axis shows the count rate by eROSITA. Here we apply the {\sc XSPEC} hydrogen atmosphere model {\sc nsatmos}. \label{fig:logNlogS}}
\end{figure}

We perform several simulations each including $N_\mathrm{HOFNAR}=2250$ objects according to the model described above. Note, that this leads to some statistical fluctuations in the results presented below, especially for bright sources,  {  as analysed objects have a new random spatial distribution in each run}. 

We present our final results in the form of Log $N$ -- Log $S$ distributions  for different surface temperatures, Fig.~\ref{fig:logNlogS}. 
Four values of  {  the surface temperature} of HOFNARs are used: $5 \times 10^5$ K, $8 \times 10^5$ K, $10^6$~K, and $1.2 \times 10^6$~K.
Obviously, for larger temperatures  eROSITA can detect more sources. {   The distributions for all temperatures are averaged over 10 runs. }

The Log $N$ -- Log $S$ lines have a slope slightly flatter than -1. This is expected, as the sources are distributed in the disc, and at larger distances interstellar absorption becomes important. At the bright end the lines go steeper, asymptotically approaching the slope -3/2 for the brightest (and so -- the closest) sources. 

For the fiducial value $10^6$~K about 500 sources are predicted above the detection limit 0.01 cts~s$^{-1}$. 
However, the identification of dim sources might be a very difficult task. Thus, the number of sources with count rate  $> 0.1$ cts~s$^{-1}$ might be a better reference point. 
{  Even for the minimal exposure $\gtrsim 1600$~s accounting for the vignetting coefficient 1.88 in four years of the survey this might provide $\gtrsim 85$ photons which is enough for rough spectral classification (e.g., to identify the thermal spectrum and estimate the temperature). Also we remind, that the survey was terminated early in 2022 and at the moment, it is unclear if the whole 4-year program can be completed. So, in addition to results on barely detectable sources we also separately discuss sources with CR$>0.1$~cts~s$^{-1}$, sometimes focusing particularly on them. For the surface temperature $10^6$~K the number of such sources is $\sim 100$.} 


For surface temperatures $\lesssim 5 \times 10^5$ K the number of sources becomes too low to hope for successful identifications, {  unless the actual ratio of the number of HOFNARs to the number of MSPs is very close to the upper limit discussed above. For even lower temperatures perspectives to detect objects with the eROSITA all-sky survey becomes illusive.} 

\begin{figure}[t]

    \includegraphics[width=\textwidth]{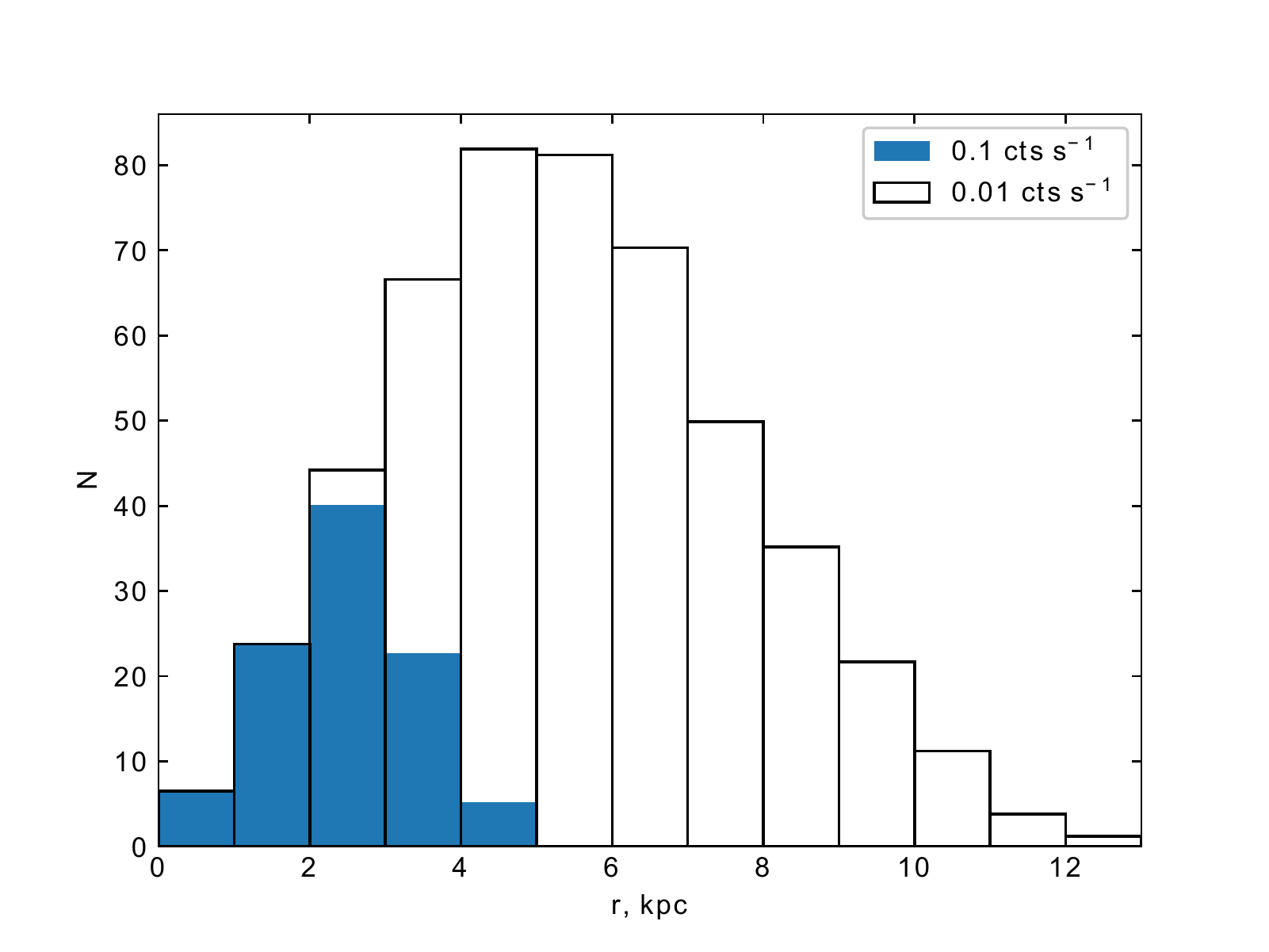}

    \caption{Radial distributions {  (distance from the Sun)} of sources with count rates 
    $> 0.01$ cts~s$^{-1}$ {  (unfilled histogram) and  $> 0.1$ cts~s$^{-1}$ (filled histogram)} for $T=10^6$ K. {   The distributions are averaged over 10 runs. }
    \label{fig:radial}}
\end{figure}

In Fig.~\ref{fig:radial} we show {  two} distance distributions of HOFNARs with $T=10^6$ K and count rates 
    $> 0.1$ cts~s$^{-1}$ and {  $> 0.01$ cts~s$^{-1}$}. It can be seen, that a typical distance {  from the Sun} is {  $\sim$3--8 kpc  for the dimmer sources and $\lesssim$ 4--5 kpc -- for the brighter}.
    
\begin{figure}[t]
\begin{adjustwidth}{-\extralength}{0cm}
\centering
    \includegraphics[width=\textwidth]{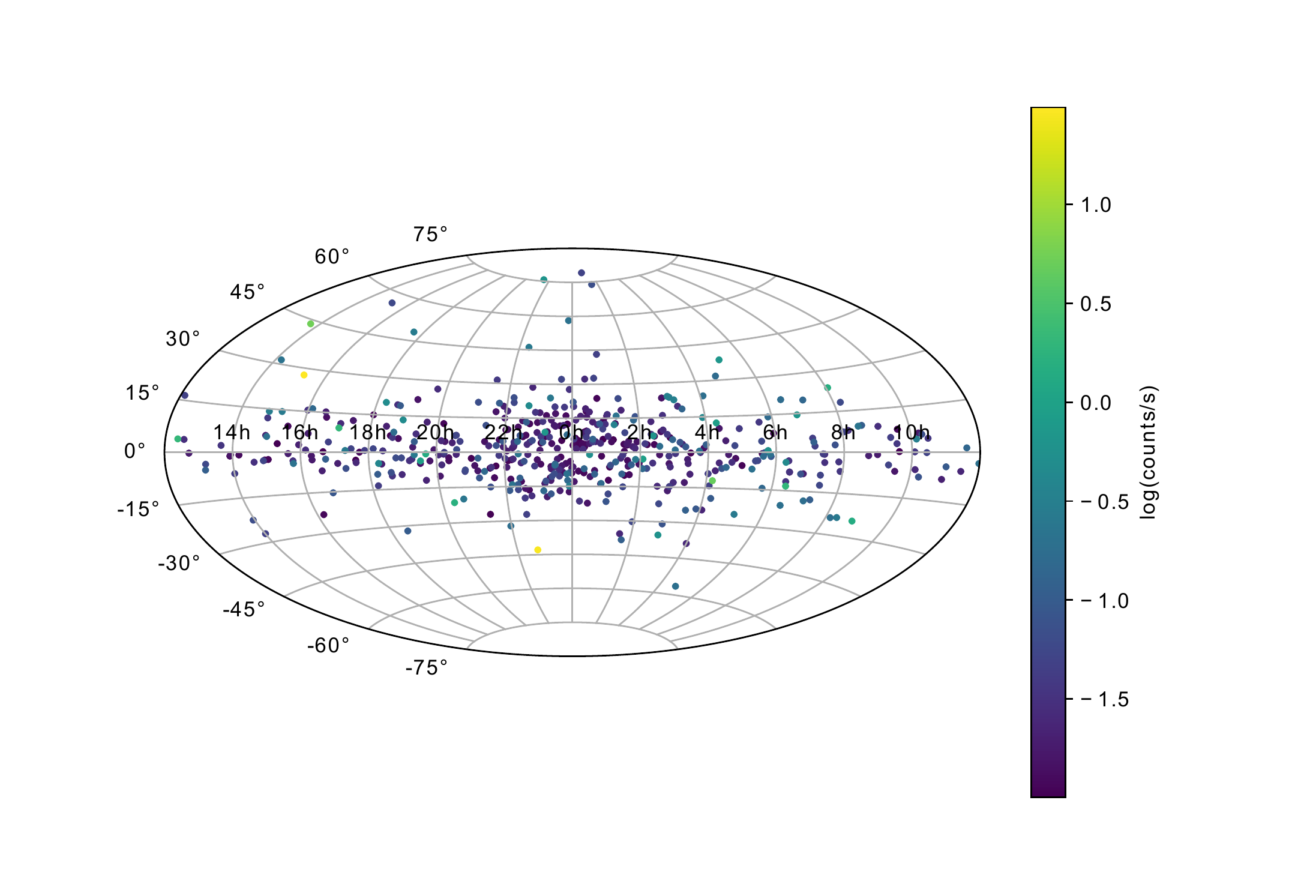}
\end{adjustwidth}
    \caption{Map of sources with count rates 
    $> 0.01$ cts~s$^{-1}$ 
    in the Galactic coordinates {  for a single run with a total of 2250 sources}. The surface temperature is $T=10^6$ K. Color of each circle corresponds to the count rate (see the bar on the right). 
    \label{fig:map}}
\end{figure}

    Distribution in the sky is shown in Fig.~\ref{fig:map}. As expected, bright near-by sources {  (lighter symbols)} can be found at large Galactic latitudes, where it can be easier to identify them, but such objects are not numerous.


{  Concluding presentations of our main results, we have to say that uncertainties in the absolute numbers of sources can reach a factor $\lesssim (2-3)$ due to poor knowledge of several parameters, in the first place -- the relative number of HOFNARs and MSPs. }

\section{Discussion} 
\label{Sec_Discus}

 Population synthesis calculations presented above demonstrate that for realistic parameters HOFNARs can be detected in the all-sky survey by eROSITA. Still, it is appropriate to discuss several issues related to non-detections in earlier {  surveys} and difficulties in identifying HOFNARs among dim eROSITA sources. 

\subsection{Comparison with ROSAT} \label{Sec_ROSAT}

Before eROSITA, the best all-sky survey in soft X-rays was done by ROSAT. The final catalogue (2RXS) includes $\sim135$~000 items  \cite{2016A&A...588A.103B}. Significant fraction of these detections can be spurious, many sources are unidentified. 
The situation is much better for the bright source catalogue -- RASS-BSC \cite{1999A&A...349..389V}. In this sample there are $\sim18000$ sources with count rate $>0.05$~cts~s$^{-1}$. 
A very high rate of identification is achieved in the 
ROSAT Bright Survey (RBS). This catalogue includes objects with count rate $> 0.2$ counts~s$^{-1}$ and Galactic latitude $|b| > 30^{\circ}$ \cite{2000AN....321....1S}. It is useful to compare our model with this sample in order to be sure that we do not over predict the number of bright sources.

We use our model and data on the response function of ROSAT (\cite{1987SPIE..733..519P}, see Fig.~\ref{fig:Seff}) to calculate the number of HOFNARs and their distribution in the sky.
The number of sources with ROSAT counts $> 0.2$ counts~s$^{-1}$ and Galactic latitude $|{b}| > 30^{\circ}$ for $T = 8 \times 10^5$ K on average is $\sim1$ (the range is 0~--~3 in $\sim 10$ 
realizations).
For $T = 1 \times 10^6$~K we obtain 3 sources (the range 2~--~4),
while for $T = 1.2 \times 10^6$ K  predicted number of detected sources becomes $\sim 6$ (the range 3~--~10). 
In the last case the predicted number of bright sources for ROSAT {  starts to be} too
big, as no objects which can be HOFNARs are identified in the RBS. This provides an opportunity to set an upper limit for the temperature of HOFNARs 
{  for our fiducial $\nu$}. Results for temperatures $\lesssim 10^6$~K do not contradict non-detection of HOFNAR candidates by ROSAT {  on the $\sim 2\sigma$ level,  suggesting that our fiducial choice $\nu=0.05$ is, in fact, an upper limit imposed by ROSAT for the $T=10^6$~K}. 
{  
As in the case of predictions for detectability of HOFNARs by eROSITA, in calculations for the ROSAT observations this coefficient is the most uncertain quantity in our calculations. This uncertainty, which can easily  reach a factor $\sim2-3$ (or even more), excludes any more precise limitation.}

It is worth to note, that the fact that ROSAT discovered seven  X-ray Dim Isolated Neutron Stars (XDINSs)
known as the Magnificent Seven \cite{Haberl07, 2009ASSL..357..141T},  but not a single HOFNAR candidate  should not be treated {  automatically} as an argument that eROSITA should discover {  a much} larger number of XDINSs than HOFNARs.
The Magnificent Seven {  has its} origin in the Gould Belt --- the local   {  ring-like structure made} of young stars which encompasses the Sun \cite{Popov_ea03}. So, the number density of young isolated NSs is enhanced in the Solar vicinity {  by a factor $\lesssim 3$} and does not represent the average number density in the Galactic disc.
{  At larger distances in favourable conditions ($\nu\gtrsim0.05$, $T\sim 10^6$~K) HOFNARs can be found in comparable number, or even outnumber XDINSs. Distinguishing between these two populations can be a difficult task. We briefly discuss some issues in the following subsection.}

\subsection{On identification of HOFNARs in the X-ray survey}

As shown in Sec.~3, for our fiducial parameters (see Sec.\ \ref{Sec_Model}) we expect detection of {  $\sim 500$} HOFNARs in the completed eROSITA all-sky survey and {  $\sim 100$} of them (i.e., {  $\lesssim$ 5\% } of the total HOFNAR population {  in the Galaxy}) are expected to have rather high count rate $>0.1$~s$^{-1}$ leading to a significant number of photons for {  rough} spectral analysis.
However, it is crucial not only to detect, but also to identify a source as a HOFNAR.
Here we discuss some HOFNAR features, which can be 
used for
identification. 
However, a detailed analysis of the identification strategy is beyond the scope of this paper. 

Identification of sources of a given type can be difficult if there is another class of sources with very similar properties.
In the case of HOFNARs the main source of confusion can be due to  young isolated NSs, which might be detected  by eROSITA in comparable number \cite{2017AN....338..213P, 2022arXiv220107639K},
and also due to qLMXBs.

Let us start with isolated NSs.
The analysis of detectability of isolated NS with non-thermal radiation by eROSITA  was recently carried out in  \cite{2022arXiv220107639K}. It was found that among known sources eROSITA can detect $\sim 160$ radio pulsars and  $\sim 20$ magnetars. 
Some isolated NSs -- XDINSs and central compact objects (CCOs), -- do not demonstrate non-thermal components in their spectra.
It is expected that eROSITA will detect all known XDINSs and most of CCOs. New discoveries are also expected.
According to \cite{2017AN....338..213P}, eROSITA will be able to detect $\gtrsim 80$ new XDINSs.

Hopefully, 
purely thermal X-ray emission allows to discriminate HOFNARs from significant part of isolated NSs as many of them have non-thermal components in their spectra.
A power law component is typically required to describe a high energy tail in the X-ray emission of highly magnetized  NSs (i.e. magnetars, see e.g., \cite{ok14_McGill_Catalog}) as well as many of NSs with ordinary magnetic fields (see e.g., \cite{Potekhin_ea20_ThermLum} and references therein).
It is also easy to identify CCOs (see Ref.\ \cite{DeLuca17} for a review) as they are located inside supernova remnants which is not expected for very old HOFNARs.
Still, XDINSs
require additional criteria to be discriminated from HOFNARs.%

 X-ray pulsations can be the first of such criteria.
HOFNARs should have almost isothermal surface and their emission should not pulsate or pulsate at the spin rate, which should be high enough to  allow the r-mode instability  ($\gtrsim 200$~Hz) \cite{cgk14}. Thus, detection of low frequency pulsations excludes identification of a source as a HOFNAR.
{  We remind that} pulsations of  X-ray flux with periods 3-12 s are detected for at least six out of seven classical XDINSs \cite{2009ASSL..357..141T}. {  Still, newly discovered XDINSs are expected to be dim objects with low count rate, even for eROSITA (in the survey mode). So, before they are observed with long exposures by X-ray observatories like XMM-Newton, Chandra, eROSITA (in the pointing observation mode), and (in near future) ATHENA  it might be impossible to detect pulsations.}

The second criterion can be based on a detailed analysis of X-ray spectra including their long term variability. HOFNARs should be stable sources and no absorption features are predicted by hydrogen and helium atmosphere models for HOFNAR temperatures in the eROSITA energy range, see e.g. \cite{zps96}. This is in contrast with properties of the Magnificent Seven. So, presence of such features in the data  excludes the HOFNAR interpretation. It is crucial that in principle, this analysis can be done solely on the base of eROSITA data {  if enough photons are collected} (see, e.g.\ \cite{psk22} for analysis of some of the known XDINS), however contribution from other X-ray observatories is welcomed, as well as long dedicated eROSITA observations after the survey is completed.

An additional criterion of HOFNAR identification can be associated with X-ray polarization.\footnote{Imaging X-Ray Polarimetry Explorer mission -- IXPE, --  was recently launched \cite{IXPE}. Effects of the magnetized vacuum were predicted to affect X-ray emission of XDINSs, e.g. \cite{gczttw16}.}   {  However, detection of polarization from dim X-ray sources is a challenging task. Future missions like eXTP \cite{2019SCPMA..6229502Z} potentially can succeed here. }

Some other criteria are the following: absence of  radio and  gamma-ray counterparts demonstrating pulsations at non-millisecond periods, position of a source in the Galaxy (young NSs should be located in the regions with active star formations, while HOFNARs can be located at large Galactic latitudes, see Fig.\ \ref{fig:map}). 

Finally, HOFNARs are born in LMXBs and the secondary component can be not exhausted {  completely}, yet. In this case, it can reveal itself in optical emission. 
For example,  the X-ray source X5 in 47 Tucanae, which is a HOFNAR candidate \cite{cgk14},
have a faint optical counterpart  ($V=21.7$, $U-V=0.9$) which can be interpreted as a companion not filling the Roche lobe \cite{2002ApJ...564L..17E}. 
Taking into account the distance to 47 Tuc  (4.45 kpc) we can expect that if a HOFNAR has a secondary component with similar properties, then on average it should be detected  (see Fig.\ \ref{fig:radial} for the expected distance distribution for detectable HOFNARs). 
Oppositely, even the brightest source among the Magnificent Seven -- RX J1856.5-3754, -- at the distance $\sim 140$~pc has $V\sim 25.6$ \citep{2001A&A...378..986V, 2007MNRAS.375..821H}, being clearly undetectable
if located at $\sim2-4$~kpc distance. 
Correspondingly, many faint isolated cooling  NS candidates (see, e.g. \cite{2022MNRAS.509.1217R}) might not have optical counterparts. For example, the authors of \cite{2017AN....338..213P} predict that just $\sim 25$ among the expected $\sim 80$ new XDINS (to be discovered by eROSITA) might have optical counterparts, {  which are expected to be very dim}.  

It can be extremely hard to discriminate HOFNARs from hot qLMXBs, because the main difference between these objects is the state of a companion star: for a qLMXB the companion should fill the Roche lobe to allow (transient) accretion, required to maintain the NS temperature (e.g., \cite{pcc19} and references therein). Even known sources in globular clusters are still considered as candidates to both classes (see \cite{cgk14} for discussion). 
Likely, important constraints for the qLMXB population potentially discovered by eROSITA
can be obtained combining observational data on LMXBs with predictions of the disc instability model \cite{Lasota01}
(see \cite{cgk14}  for a similar analysis for sources in globular clusters), but we leave these problems beyond this paper.

Also, there can be statistical features supporting the hypothesis that the HOFNAR population is 
indeed detected. For example, if the temperature distribution of soft X-ray sources is clustered 
in
a very narrow temperature range -- it would be natural for HOFNARs, but not for isolated NSs which 
continuously cool down and should have a smoother temperature distribution.
In the absence of resonance stabilization of r-modes \cite{gck14a,gck14a}, temperatures of qLMXBs 
are determined by an average accretion rate  (e.g., \cite{pcc19}) and also should have a smooth 
distribution.
However, a practical application of this criterion requires a careful  analysis of  selection 
effects which clearly affects the observed population.
Importance of such analysis, 
which is, however, beyond the scope of the present paper, 
is highlighted by the fact that temperatures of six XDINSs discovered by ROSAT 
lie
in {  a relatively} narrow 
range $85-102$~eV \cite{Haberl07}. 

Clearly, detection and identification of HOFNARs would be a magnificent result as they represent a new class of NSs with a specific source of thermal energy.
It would not just confirm that the r-mode instability can be active in NSs, but also that it plays an important role in observational appearance of some LMXBs.
Still, let us consider a possible negative case: none of X-ray sources discovered by eROSITA can be identified as a HOFNAR.
It would be a clear indication that at least one of the assumptions used in our model (Sec.\ \ref{Sec_Model}) is wrong.

The most direct interpretation of the negative result is the following: the number of Galactic HOFNARs  with temperatures about the fiducial value $10^6$~K is very small.
As long as $\lesssim 5$\% of HOFNARs  should be detected above 0.1 cts~s$^{-1}$, a non-detection might limit the total Galactic number of HOFNARs with
$T\gtrsim  10^6$~K by  $\lesssim $ {  a few tens}.
This can have two ({  not mutually exclusive}) 
explanations: either the probability of the birth of a HOFNAR in a LMXB is much lower than our fiducial value,%
\footnote{Here we treat as a newborn HOFNAR a NS which finishes the LMXB stage of evolution with active r-mode instability.}
or/and the life time of a HOFNAR at the temperature not much lower than the fiducial one is much smaller than the Galactic age.
For example, the latter can be the case if LMXB evolution can lead to formation of heavy  HOFNARs, with masses higher than the direct URCA threshold. In this case, 
HOFNARs would have a huge cooling power $L_\mathrm{cool}$ due to high neutrino luminosity. As far as HOFNAR evolution is governed by the following equation, see \cite{cgk14}
\begin{equation}
    \frac{\mathrm d E_\mathrm{rot}}{\mathrm d t}=-3 L_\mathrm{cool},
    \label{Evolution}
\end{equation}
this would result in a short life time of a bright source. Here $E_\mathrm{rot}$ is the rotational energy of a HOFNAR and $t$ is time.
Let us also remind that LMXB evolution in globular clusters is more complicated due to close encounters and, in principle, this can lead to a more efficient production of observable HOFNARs than in the Galactic disc.

%
 

\section{Conclusions} \label{Sec_Concls}

 In this paper we presented a population model of HOFNARs -- NSs originated in LMXBs and heated due to the r-mode instability. We demonstrate that in the framework of our fiducial model there are significant chances that from few tens up to several hundred of these sources will be detected in the eROSITA all-sky survey. 
 If not -- then, important constraints can be put on properties and origin of such objects. 

\vspace{6pt} 


\authorcontributions{A.C. initiated the study and supervised aspects of the model related to thermal properties of HOFNARs. A.K. and S.P. made the population synthesis model. The code was written by A.K., she also made calculations. E.K. and M.G. contributed to the discussion of results.
M.G. has checked the calculations by independently written code.
}

\funding{A.K. and S.P. are supported by the Russian Science Foundation, grant 21-12-00141.
}




\dataavailability{ Observational data used in this paper are quoted from the
cited works. Data generated from computations are reported
in the body of the paper. Additional data can be made available upon reasonable request.
} 

\acknowledgments{
We thank the referees for useful comments. 
}

\conflictsofinterest{The authors declare no conflict of interest.} 


\abbreviations{Abbreviations}{
The following abbreviations are used in this manuscript:\\

\noindent 
\begin{tabular}{@{}ll}
CCO & Central compact object\\
eROSITA & extended ROentgen Survey with an Imaging Telescope Array\\
HOFNAR & HOt and Fast Non Accreting Rotator\\
LXMB  & Low-mass X-ray binary\\
MSP & Millisecond radio pulsar\\
NS & Neutron star\\
qLXMB  & quiescent Low-mass X-ray binary\\
SRG & Spectrum-Roentgen-Gamma
\end{tabular}}




\begin{adjustwidth}{-\extralength}{0cm}

\reftitle{References}


\bibliography{bib}

\end{adjustwidth}
\end{document}